# Wide-Field Mueller Polarimetry of Brain Tissue Sections for Visualization of White Matter Fiber Tracts

PHILIPPE SCHUCHT,[1,5,†] HEE RYUNG LEE,[2,†] MOHAMMED HACHEM MEZOUAR,[2] EKKEHARD HEWER,[3] ANDREAS RAABE,[1] MICHAEL MUREK,[1] IRENA ZUBAK,[1] JOHANNES GOLDBERG,[1] ENIKÖ KÖVARI,[4] ANGELO PIERANGELO[2,6], AND TATIANA NOVIKOVA[2,7]

[1]*Department of Neurosurgery, Inselspital, Bern University Hospital, and University of Bern, Bern 3010, Switzerland*
[2]*LPICM, CNRS, Ecole polytechnique, Institut Polytechnique de Paris, Palaiseau, 91128, France*
[3]*Department of Pathology, Inselspital, Bern University Hospital, and University of Bern, Bern 3010, Switzerland*
[4]*Department of Mental Health and Psychiatry, University Hospitals of Geneva, Geneva, 1205, Switzerland*
[5]*philippe.schucht@insel.ch*
[6]*angelo.pierangelo@polytechnique.edu*
[7]*tatiana.novikova@polytechnique.edu*
[†]*these authors contributed equally: Philippe Schucht, Hee Ryung Lee*

**Abstract:** Identification of white matter fiber tracts of the brain is crucial for delineating the tumor border during neurosurgery. A custom-built Mueller polarimeter was used in reflection configuration for the wide-field imaging of thick section of fixed human brain and fresh calf brain. The experimental images of azimuth of the fast optical axis of linear birefringent medium showed a strong correlation with the silver-stained sample histology image, which is the gold standard for ex-vivo brain fiber tract visualization. The polarimetric images of fresh calf brain tissue demonstrated the same trends in depolarization, scalar retardance and azimuth of the fast optical axis as seen in fixed human brain tissue. Thus, label-free imaging Mueller polarimetry shows promise as an efficient intra-operative modality for the visualization of healthy brain white matter fiber tracts, which could improve the accuracy of tumor border detection and, ultimately, patient outcomes.

## 1. Introduction

Surgery is the crucial treatment step for most patients with brain tumors [1-7]. Clear identification of the exact border between the tumor and the surrounding brain tissue is essential to allow radical tumor resection and to preserve neurological function. However, although it is easy to identify the tumor in preoperative magnetic resonance imaging (MRI), solid tumor tissue is often difficult to differentiate from infiltrated white matter during surgery, even when using a state-of-the-art intra-operative microscope.

Patients in whom a piece of the tumor is left behind due to poor visualization of the tumor border have worse prognosis than those in whom the entire tumor was removed, as the tumor invariably grows back from the remnants [8-17]. Furthermore, information on the neurological function of a given area of exposed white matter seen during surgery is very limited. The white matter of the healthy brain is made up of fiber tracts that comprise bundles of axons. This highly ordered structure of healthy white matter is very different from brain tumor tissue, whose cells grow in a largely chaotic way (see Fig. 1). However, this difference in structural complexity is currently not detectable during surgery with a white-light surgical microscope. These difficulties in identifying tumor, function and fiber tracts are key contributors to the risk of both incomplete resection (too little resection) and neurological deficits (too much resection).

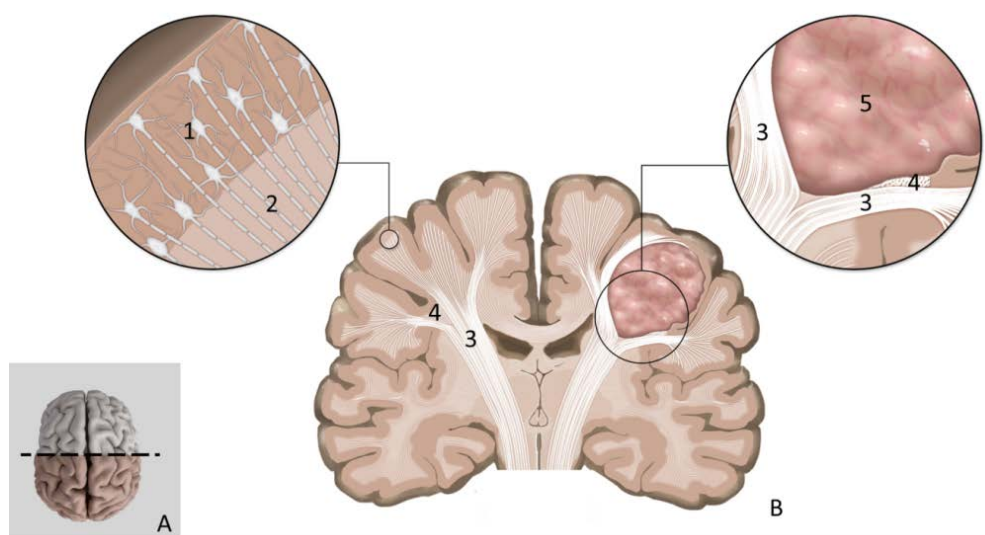

Fig. 1. Schematic architecture of brain: A – view of a brain from above, the dashed line shows the location of a coronal plane; B – schematic of a brain cross-section in a coronal plane. Left inset: 1 – cell bodies of neurons lie within the brain gray matter; 2 – axons; the neurons connect to other areas of the brain or to the spinal cord via their axons. Right inset: 3 – corticospinal tract; 4 – anterio-posterior-running arcuate fasciculus fiber tract; 5 – tumor. See explanations in the text.

So far several experimental methods have been investigated for their ability to discern brain tumor tissue. For instance, orally administered 5-aminolevulinic acid (5-ALA) is taken up by cells and metabolized to protoporphyrin IX, which accumulates in tumor cells of higher-grade gliomas and exhibits fluorescent properties [18-21]. By illuminating the exposed brain tissue with blue light through the surgical microscope, tumor cells fluoresce red, making tumor cell clusters visible. 5-ALA has entered clinical routine [18, 22, 23], however, its benefit is limited to high-grade gliomas, as it is not suitable for visualizing low-grade gliomas or metastases due to insufficient protoporphyrin IX accumulation.

The installation of an MRI device in the operating room helps to identify tumor remnants during surgery and has been shown to increase the rate of gross total resection in patients with high-grade glioma [25-27]. However, the significant financial costs and extra time needed for scanning prevent intra-operative MRI from becoming standard of care. Other approaches, such as intra-operative ultrasound, have not proven reliable in estimating the extent of resection and the residual tumor volume [28]. In summary, the attempts to visualize tumor cells have so far failed to reliably identify the tumor–brain interface during surgery for many intrinsic brain tumors.

The schematic of a brain section in a coronal plane (see Fig. 1A) is shown in Fig.1B. The cell bodies of neurons lie within the superficial layer of the brain, which is called the gray matter or cortex (see Fig.1B, left inset). The axons, constituting the white matter of the brain, conduct electrical impulses (action potential) between nerve cell bodies located in the gray matter of the brain or in the spinal cord. Large numbers of axons are joined together in fiber tracts. For example, the corticospinal tract connects to the spinal cord and is responsible for voluntary movement of the limbs, whereas the anterio-posterior-running arcuate fasciculus fiber tract is responsible for speech (see Fig.1B, right inset). A brain tumor displaces these fiber tracts.

Each axon is surrounded by a myelin sheath, which acts as an electrical insulator to accelerate propagation of action potential. Myelin is a lipid-rich substance with a refractive index higher than that of the surrounding glia, in the visible wavelength range [29]. The white color of the inside of the brain is due to strong scattering of light, which results in depolarization of incident polarized light. In addition to light scattering, the densely packed and aligned rods of myelin produce strong optical anisotropy (so-called "form birefringence") of brain white matter. Consequently, brain fiber tracts must exhibit uniaxial linear birefringence with the optical axis oriented along the direction of the fiber bundle. The speed of light propagation through healthy brain tissue will depend on the direction of propagation, resulting in different phase shifts of electric field components of plane polarized electromagnetic wave propagating through a birefringent medium.

Earlier polarimetric studies of a thin histological section of brain tissue in transmission configuration have confirmed that fiber bundles in white matter cause optical anisotropy of brain tissue [30, 31]. The visualization of brain fiber tracts with polarized light was demonstrated in reflection configuration at microscopic image resolution [32]. The phenomenon of contrast enhancement between cancerous and healthy zones of tissue on wide-field images of depolarization, scalar linear retardance, and azimuth of optical axis measured in backscattering configuration has already been established for thick specimens of fresh human colon, rectum, and uterine cervix, as well as for in vivo uterine cervix [33-37]. This contrast enhancement was explained by the disruption of the extracellular collagen matrix in malignant tissue, which leads to the loss of optical anisotropy.

Brain tumors also destroy the highly ordered structure that is characteristic of healthy white matter and erase the optical anisotropy of healthy brain. Instead of focusing on detecting the tumor itself, we suggest an alternative approach, namely, detecting healthy white matter by means of its fiber tracts. Here we present for the first time to our knowledge the results of using wide-field imaging Mueller polarimetric system operating in the visible wavelength range for the visualization of the direction of fiber tracts of white matter of healthy brain tissue. Measuring Mueller matrix images of thick sections of brain tissue and using a non-linear data post-processing algorithm we showed that the resulting maps of azimuth of the optical axis provide rapidly and consistently the presence and the directions of fiber tracts in the imaging plane.

## 2. Materials and Methods

### 2.1 Instrument and measurement protocol

We studied unstained thick sections of both fixed human brain and fresh animal brain with wide-field imaging Mueller polarimetry in reflection, which is the most relevant configuration for clinical applications of optical techniques using visible light. A photo and schematic of ferroelectric liquid crystal-based imaging polarimeter is shown in Supplementary Materials (Fig. S1). This instrument has already been described in previous publications [38, 39]. For the sake of completeness we outline here the main characteristics and operational principles.

A linear polarizer and two voltage-driven ferroelectric liquid crystals assembled sequentially (polarization state generator – PSG) and another set of the same optical elements, but assembled in a reverse order (polarization state analyzer – PSA), were introduced into the illumination and detection arms of a conventional imaging system, respectively. A xenon lamp was used as the incoherent white light source for sample illumination. Each ferroelectric liquid crystal works as a wave plate with the fixed retardation and fast optical axis orientation switching between 0° and 45°. The PSG modulates the polarization of the incident light beam illuminating the sample at an incidence angle of about 15° and spot size ~10 cm along the main ellipse axis. The light backscattered by a sample passes through the PSA before being detected by a 512 × 386 pixels CCD camera (Stingray F080B, Allied Vision, Germany) with its optical axis placed normal to the sample imaging plane. To measure 16 elements of Mueller matrix [40] four different polarization states of incident light sequentially generated by the PSG are projected on four polarization configurations by the PSA. A rotating wheel, placed in front of the PSA, contains reference samples for calibration of the instrument, namely, two polarizers with eigenaxes oriented at 0° and 90° and a wave plate with the optical axis oriented at 30°. The optimal PSG and PSA polarization states are defined by the automated calibration procedure described previously [41, 42].

A rotating wheel, placed behind the PSA, holds the 40 nm bandpass interference filters for performing multi-wavelength measurements from 450 to 700 nm in steps of 50 nm. There are two main reasons for using the broad band filter. First, the intensity of monochromatic normally backscattered light decreases significantly, because light propagation through optically thick biological tissue is dominated by multiple scattering. Using a broadband dichroic filter helps to increase the signal-to-noise ratio of the backscattered signal. Second, the intensities at different wavelengths are summed incoherently, thus erasing the speckle patterns seen when using a coherent light source.

The measurement protocol includes the sequential acquisition of 16 intensity images for four different input and four different output polarization states at each measurement wavelength. Rapid polarization modulation supported by electrically switchable ferroelectric liquid crystals results in the acquisition of 16 images in a few seconds. Mueller matrix images of a sample are then calculated from the raw intensity measurements at each measurement wavelength [42].

The thick section of human brain tissue was removed from the formalin and placed flat in a glass Petri dish 14.5 cm in diameter. The measurements of the Mueller matrix were performed first on the tissue exposed to air, then a sufficient amount of distilled water was poured into the Petri dish to cover the surface of the tissue and optical measurements were repeated. Covering the tissue with water leads to partial index matching and flattening of the interface, which, in turn, mitigates the artifacts related to sample surface topography.

A thick section of fresh cadaveric calf brain tissue was put into the glass Petri dish and imaged with the Mueller polarimeter immediately after preparation. The polarimetric measurements were performed with calf brain tissue exposed to air, because these conditions are closest to those in clinical settings during neurosurgery.

To interpret the measured Mueller matrices from brain tissue in terms of basic polarimetric properties (depolarization, retardance, dichroism) we applied the Lu-Chipman algorithm of non-linear data compression [43], which allows decomposition of any physically realizable Mueller matrix into the product of three Mueller matrices of the basic optical elements, namely, diattenuator, retarder, and depolarizer. Lu-Chipman decomposition provides the data on the sample depolarization power, and the vectors of retardation and diattenuation.

*2.2 Sample preparation*

Human brain tissue was obtained from the autopsy of an anonymous donor. The brain was formalin-fixed, and one half of a 1 cm thick section of fixed human brain in a coronal plane (see Fig. 1A) was used for optical measurements. The dimensions of the brain section were approximately $6 \times 9$ cm2.

The remaining fixed brain tissue was then paraffin-embedded according to standard procedure. Thin whole-mount sections were prepared from the part of the brain adjacent to the part that had been imaged. Whole-mount sections were stained with Bielschowsky silver impregnation and subsequently digitized on an M8 robotic microscope (Precipoint, Freising, Germany). A waiver for ethical approval was obtained from the Ethics Committee of the Canton of Bern (KEK 2017-1189).

The whole cadaveric fresh (not fixed with formalin) calf brain was bought from a local French butcher. The brain tissue was cut in a coronal plane to prepare a ~1 cm thick section of tissue. This section of calf brain tissue was rinsed with cold water to remove visible blood clots, which can affect polarimetric measurements. Color photos of both samples are shown in Supplementary Materials (Fig. S2).

## 3. Results and Discussion

### *3.1 Fixed human brain tissue*

We explored the sensitivity of polarized light to optical anisotropy of brain white matter. The Mueller imaging system provides 16 polarimetric images of the Mueller matrix for thick sections of fixed and fresh brain tissue at each chosen measurement wavelength. The maps of basic optical properties of the sample, namely, depolarization power, retardance, and dichroism were obtained by applying pixel-wise the Lu-Chipman polar decomposition. The images of total depolarization, scalar linear retardance, and azimuth of optical axis calculated from the Mueller matrix of a thick section of fixed non-tumorous human brain tissue exposed to air and measured at 550 nm are shown in Figs 2A-2D. Neither circular birefringence nor linear or circular dichroism was detected experimentally. These are quite typical polarimetric responses of thick tissue blocks measured in reflection [33, 34, 35].

The gray-scale intensity image and image of total depolarization of the thick section of fixed brain tissue measured at 550 nm are presented in Figs 2A and 2B. The depth of light penetration in the tissue is only a few hundred microns at this wavelength. The zones of specular reflection are characterized by lower depolarization, the traces of the cutting blade on the brain specimen's surface show more contrast compared to the intensity image because they are more depolarizing. In these measurements, the depolarization response of the sample bulk is strongly affected by the surface contribution. The cortex of the brain is less depolarizing than the white matter.

The images of scalar linear retardance and azimuth of optical axis are shown in Figs 2C and 2D. As expected, the white matter of healthy brain tissue demonstrates measurable linear retardance of about 5° up to a maximum of 20° (Fig. 2C). The map of azimuth of optical axis (Fig. 2D) shows the orientation of fiber bundles.

The gray-scale intensity image, total depolarization, scalar linear retardance, and azimuth of optical axis images of the same sample immersed completely in water are shown in Figs 2E-2H. The measurements were all performed at a wavelength of 550 nm. The dashed regions represent the zone of specular reflection of the slightly divergent incident light beam from the air-water interface as detected by a CCD camera. The surface topography was flattened by the index matching (although incomplete). Hence, the impact of surface scattering on depolarization and scalar retardance images was largely mitigated. The contrast between the cortex and white matter of the brain was increased (see Figs 2B and 2F). The mean values of scalar linear retardance were higher for the water-immersed specimen than those for the same brain section measured in air. Remarkably, the maps of the azimuth of optical axis (Figs 2D and 2H) are very similar. It can be explained by the suppression of the surface scattering effect, which leads to a reduction of the contribution of photons with a short path length. As a consequence, we collect the photons that travel longer distances within brain tissue. Bulk scattering randomizes and erases the polarization for the majority of detected photons, but a tiny portion of the detected signal, which remains polarized, accumulates a larger phase shift (i.e. a larger value of scalar retardance). In the air-exposed brain sample, a fraction of the photons with a short path length had introduced only a small phase shift and, hence, did not affect the calculations of the azimuth of optical axis.

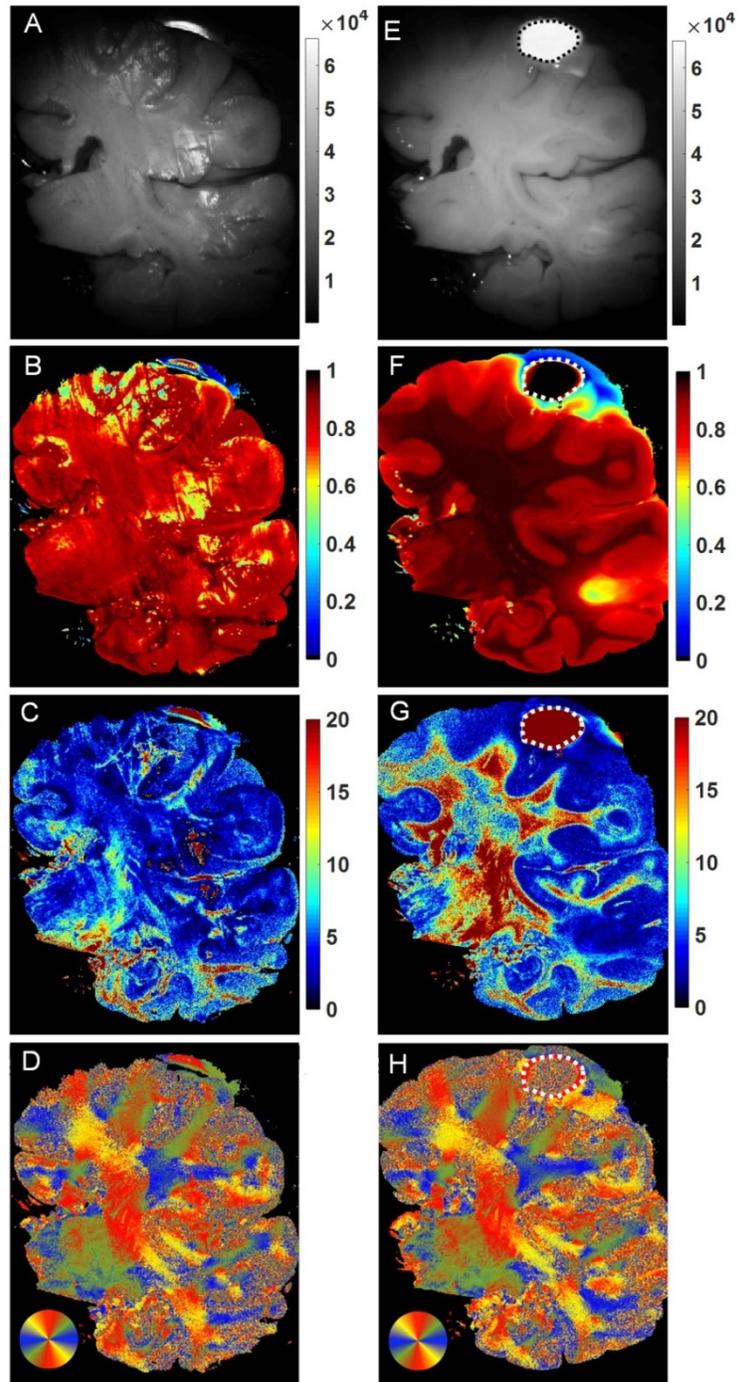

Fig. 2. 1 cm thick coronal section of a fixed human brain. Left column – images of air-exposed specimen; right column – images of water-immersed specimen; A, E – gray-scale intensity images; B, F – depolarization; C, G – scalar retardance (degrees); D, H – azimuth of optical axis. The dashed line on images E, F, G, and F delineates the area of specular reflection on the air–water interface. Note that the polarimetric images are not affected by non-uniform illumination (Figs 2A, 2E) from the slightly divergent incident light beam.

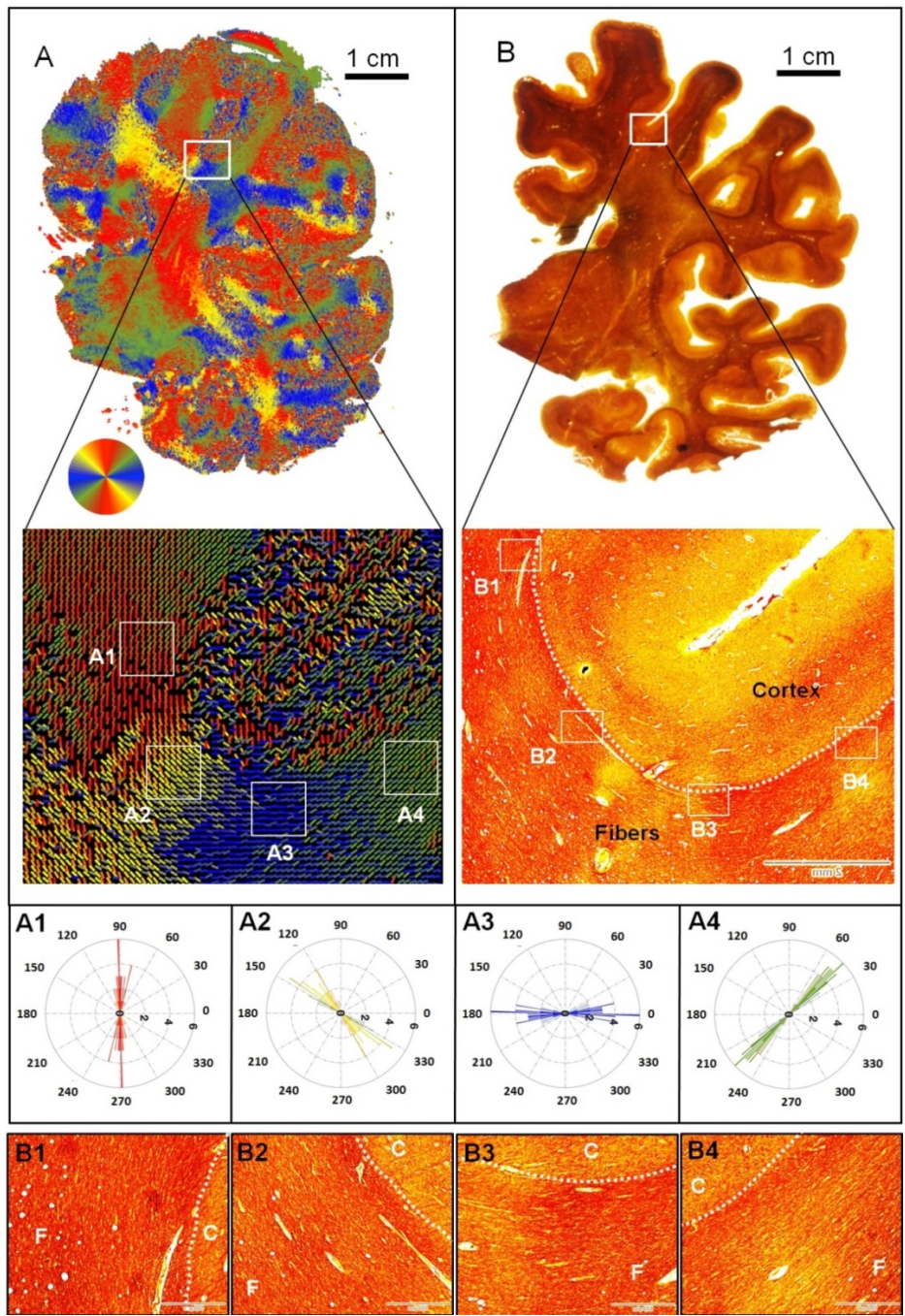

Fig. 3. Coronal section of a fixed human brain specimen: A – map of the azimuth of optical axis, bottom inset shows the enlarged zone (71 × 66 pixels) of U-fibers that curve around the superior frontal sulcus, connecting the superior and medial frontal gyrus; B – image of a silver-stained thin histological section with a bottom inset showing an enlargement of the zone depicting the same U-fibers (scale bar – 2 mm); third row – circular histograms of the azimuth of optical axis for the zones A1-A4 (9 × 9 pixels each) depicting U-fibers; bottom row – the corresponding enlarged zones B1-B4 of U-fibers on the image of the histological section (C – cortex zone, F – white matter fiber tract, white dashed line represents the border between the cortex and brain white matter, scale bar – 500 μm).

The map of azimuth of optical axis of the thick specimen of fixed brain tissue and the corresponding image of the thin, silver-stained histological section of fixed brain tissue are shown in Figs 3A and 3B. Silver-staining of thin histological section is a gold standard technique used for brain fiber visualization. To check the correlation of azimuth value with the orientation of fiber tracts, we first selected four zones of the same U-fiber tract on each image. The selected zones, A1-A4 on the azimuth map (Fig. 3A, bottom inset), represent squares of $9 \times 9$ pixels. We then calculated the circular histograms for the selected zones on the azimuth map (see Fig. 3, middle row). All circular histograms show high directionality and low spread. The mean values of the azimuth and standard deviations for the selected zones 1-4 are given in Table S1 (see Supplementary Materials).

The corresponding zones were selected on the image of the silver-stained thin histological section (see Fig. 3B). The enlarged histological images of these zones are presented in the bottom row of Fig. 3. All circular histograms demonstrate a compelling correlation with the direction of the fibers on the enlarged images of the corresponding zones of the silver-stained tissue. A set of three further zones was selected for the analysis, including one zone within the cortex and the other two at the top and bottom of the fiber T-junction in the central part of the brain specimen (see Supplementary Materials, Fig. S3). The circular histogram for the cortex zone demonstrates isotropic distribution of azimuth, thus, confirming the absence of optical anisotropy in the gray matter of the brain at the mesoscale of several hundred microns defined by the spatial resolution of our instrument (see Materials and Methods). This is supported by the enlarged histological image of this zone, which contains many cells and sparse, randomly distributed fibers. The top and bottom zones of the T-junction again demonstrate strong directionality in the circular histograms (see Table S2, Supplementary Materials) and a visible correlation with the dominant directions of fibers on the enlarged histological images of the corresponding zones.

### 3.1 Fresh animal brain tissue

To exclude the impact of tissue fixation with formalin on the polarimetric parameters, we also performed the polarimetric measurements on unfixed specimen of fresh cadaveric calf brain tissue with the same imaging Mueller polarimeter in backscattering configuration. The results of measurements at 550 nm are shown in Fig. 4.

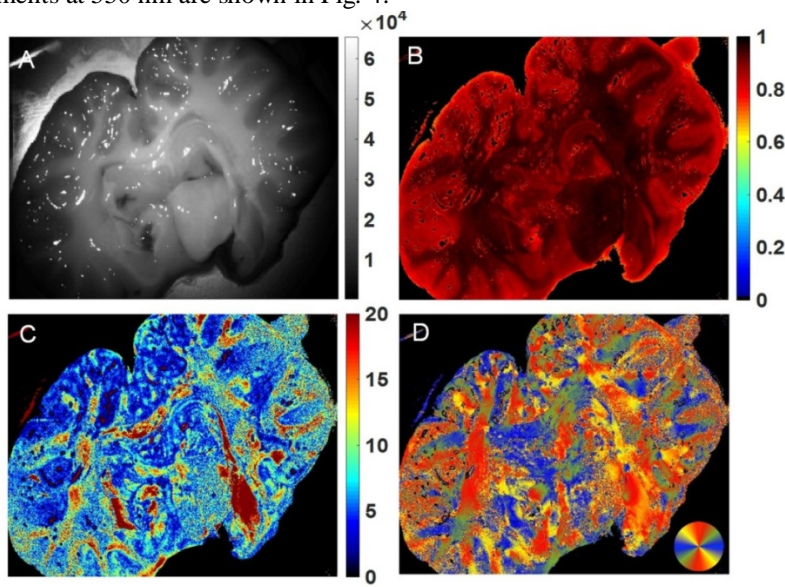

Fig. 4. Images of thick section of fresh cadaveric calf brain measured in air; A – gray-scale intensity image; B – depolarization; C – scalar linear retardance (degrees); D – azimuth of optical axis.

The bright zones on the gray-scale intensity image (Fig. 4A) are due to specular reflection because the surface of sample is not flat. The measured values of depolarization of fresh brain white matter exposed to air are comparable with the corresponding values of depolarization for the formalin-fixed tissue (Figs 2B and 4B). The contrast between gray matter and white matter on the depolarization map of fresh brain tissue is less marked than that seen in fixed tissue. It is known that tissue fixation with formalin links soluble and structural proteins together and affects optical properties of tissue, like depolarization power and scalar retardation [44]. The calculated values of scalar linear retardance for the fresh brain tissue exposed to air are comparable with the corresponding values measured for the fixed tissue immersed in water (Figs 2G and 4C). The map of azimuth of optical axis (Fig. 4D) also clearly highlights the directions of fiber tracts similar to the azimuth maps of fixed brain tissue in the white matter zone (Figs 2D and 2H). We also observe the U-fibers at the periphery of the brain specimen and thick vertical and horizontal fiber tracts intersecting in the central part of the azimuth map.

The measurements on both fixed human and fresh cadaveric animal brain tissue were also performed with wavelengths of 600 nm and 650 nm. The absolute values for depolarization and scalar linear retardance, as well as the image contrast for depolarization, scalar linear retardance, and azimuth of optical axis were very similar to those found at 550 nm. The lack of spectral sensitivity in our studies can be explained by the fact that all polarimetric measurements were made in brain tissue obtained post-mortem. There is almost no blood in a fixed tissue specimen (<1% according to the analysis done by the pathologist for the fixed brain specimen). The amount of blood in fresh cadaveric animal tissue is also significantly less than in live brain tissue and blood hemoglobin is known to be the main cause of visible light absorption in tissue, with the peaks of absorption at 500 nm and 550 nm [45]. It is highly likely that the spectral dependence of polarimetric parameters in the visible wavelength range will also reveal contrast enhancement when imaging live brain tissue. The depth of light penetration in fresh tissue depends strongly on the wavelength used [46, 47]. Taking polarimetric images at different wavelengths in real time during neurosurgery may help surgeons to estimate the remaining tumor thickness and guide tumor resection.

## 4. Conclusions

We demonstrated the feasibility of visualizing white matter fiber tracts with a wide-field imaging Mueller polarimeter operating in the visible wavelength range in backscattering configuration. The main finding of our study is the sensitivity of backscattered polarized light to optical anisotropy, induced by the densely packed neuron bundles constituting the fiber tracts of healthy brain white matter, which are not visible to the naked human eye. This result was confirmed by histological analysis of a silver-stained thin section of a brain specimen. The directions of fibers in the brain white matter, which are visible on the enlarged optical *transmission* microscopy images of a silver-stained *thin section*, are well represented by the azimuth of the optical axis calculated from the Mueller matrix images of a *thick specimen* measured in *reflection*.

Our findings open the field for the clinical implementation of Mueller polarimetry, an optical imaging technique with several key advantages. First, it is a wide-field polarimetric imaging modality, which does not require sample scanning or image stitching. This makes it faster and easier to use than polarization sensitive optical coherence tomography [48] or polarization sensitive optical coherence microscopy [32], two-photon excited fluorescence, and second harmonic generation microcopy [49]. Second, our imaging Mueller polarimeter operates in the visible wavelength range, which precludes any potential harm to patients, and is based on reflection geometry, which is a major step toward in vivo applications.

Mueller polarimetry is a non-invasive and non-contact optical technique. It does not require any sample preparation and is fast enough to be implemented in real time during neurosurgery. Our imaging polarimeter measures the complete Mueller matrix of a sample and provides the whole set of polarimetric properties (diattenuation, retardance, and depolarization). Well-developed Mueller matrix algebra [50] puts forward a variety of algorithms for the decomposition of Mueller matrices and extraction of the most valuable polarimetric parameters, for example, for increasing the image contrast [51] and improving the accuracy of medical diagnostics [38]. Finally, the design of the Mueller polarimeter can be miniaturized and the instrument can be constructed as a polarimetric add-on to a standard microscope for neurosurgery, as has already been done for a standard colposcope [37], which is used in gynecology.

We showed that wide-field Mueller polarimetry of thick sections of brain tissue in backscattering configuration clearly demonstrates the presence of fiber tracts on the images of total depolarization and linear retardance. Moreover, the orientation of the fiber tracts in the brain white matter is visualized on the map of azimuth of optical axis. This is the most robust indicator of fiber tract directions, and it works well for both fixed and fresh brain tissue. The azimuth of optical axis is almost insensitive to surface roughness, which is important for the envisaged clinical applications, where the surface morphology of surgical site will differ significantly from the conditions of our proof-of-principle studies.

The long-term aim of our studies is the creation of a miniaturized Mueller polarimeter together with optimal data processing algorithms that can be integrated into the state-of-the-art operative microscopes. This will help neurosurgeons in identifying the brain–tumor interface and identify especially important fiber tracts due to their spatial orientation, thus making brain tumor surgery both safer and more effective.

## 5. Funding and acknowledgments

### 5.1 Funding

H. R. L. gratefully acknowledges the funding from the doctoral school "Interfaces" of the École polytechnique, France.

### 5.2 Acknowledgments

H. R. L., M. H. M., A. P., and T. N. acknowledge the help of Jean-Charles Vanel with the instrument electronics. H. R. L., M. H. M., A. P., and T. N. thank Pere Roca-i-Cabarrocas and Yvan Bonnassieux for their encouragement and support of these studies. All authors acknowledge the HORAO challenge and Ivan Gusachenko for being catalysts for this collaboration.

### 5.3 Disclosures

The authors declare no conflicts of interest

## References

1. R. Soffietti, B. G. Baumert, L. Bello, A. Von Deimling, H. Duffau, M. Frénay, W. Grisold, R. Grant , F. Graus, K. Hoang-Xuan, M. Klein, B. Melin, J. Rees, T. Siegal, A. Smits, R. Stupp, W. Wick "Guidelines on management of low-grade gliomas: Report of an EFNS-EANO Task Force" Eur J Neurol. 17(9), 1124–1133 (2010)
2. C. L. Pedersen, B. Romner "Current treatment of low grade astrocytoma: A review" Clin. Neurol. Neurosurg. 115(1), 1–8 (2013)
3. E. B. Claus, P. M. Black "Survival rates and patterns of care for patients diagnosed with supratentorial low-grade gliomas: Data from the SEER program, 1973-2001" Cancer 106(6), 1358–1363 (2006)
4. P. Y. Wen, S. Kesari "Malignant gliomas in adults" N. Engl. J. Med. 359(5), 492–507 (2008)
5. A. Mintz, J. Perry, K. Spithoff, A. Chambers, N. Laperriere "Management of single brain metastasis: A practice guideline" Curr. Oncol. 14(4) 131–143 (2007)
6. R. A. Patchell, P. A.Tibbs, J. W. Walsh, R. J. Dempsey, Y. Maruyama, R. J. Kryscio, W. R. Markesbery, J. S. Macdonald, and B. Young "A randomized trial of surgery in the treatment of single metastases to the brain" N. Engl. J. Med. 322(8), 494–500 (1990)
7. A. S. Jakola, A. J. Skjulsvik, K. S. Myrmel, K. Sjåvik, G. Unsgård, S. H. Torp, K. Aaberg, T. Berg, H. Y. Dai, K. Johnsen, R. Kloster, O. Solheim "Surgical resection versus watchful waiting in low-grade gliomas" Ann. Oncol. 1942–1948 (2017)

8. N. Sanai, M.-Y. Polley, M. W. McDermott, A. T. Parsa, M. S. Berger "An extent of resection threshold for newly diagnosed glioblastomas" J Neurosurg. 115(1), 3–8 (2011)
9. M. Lacroix, D. Abi-Said, D. R. Fourney, Z. L. Gokaslan, W. Shi, F. DeMonte, F. F. Lang, I. E. McCutcheon, S. J. Hassenbusch, E. Holland, K. Hess, C. Michael, D. Miller, R. Sawaya "A multivariate analysis of 416 patients with glioblastoma multiforme: prognosis, extent of resection, and survival" J Neurosurg. 95(2), 190–198 (2001)
10. N. Sanai, M. S. Berger "Glioma extent of resection and its impact on patient outcome" Neurosurgery 62(4) 753–764 (2008)
11. M. Salvati, A. Pichierri, M. Piccirilli, G. M. Floriana Brunetto, A. D'Elia, S. Artizzu, F. Santoro, A. Arcella, F. Giangaspero, A. Frati, L. Simione, A. Santoro "Extent of tumor removal and molecular markers in cerebral glioblastoma: a combined prognostic factors study in a surgical series of 105 patients" J Neurosurg. 117(2), 204–211 (2012)
12. L. Capelle, D. Fontaine, E. Mandonnet, L. Taillandier, J. L. Golmard, L. Bauchet, J. Pallud, P. Peruzzi, M. H. Baron, M. Kujas, J. Guyotat, R. Guillevin, M. Frenay, S. Taillibert, P. Colin, V. Rigau, F. Vandenbos, C. Pinelli, H. Duffau; French Réseau d'Étude des Gliomes "Spontaneous and therapeutic prognostic factors in adult hemispheric World Health Organization Grade II gliomas: a series of 1097 cases" J Neurosurg 118(6):1157–1168 (2013)
13. M. J. McGirt, K. L. Chaichana, F. J. Attenello, J. D. Weingart, K. Than, P. C. Burger, "Extent of surgical resection is independently associated with survival in patients with hemispheric infiltrating low-grade gliomas" Neurosurgery 63(4) 700–707 (2008)
14. W. Stummer, H. J. Reulen, T. Meinel, U. Pichlmeier, W. Schumacher, J. C. Tonn, V. Rohde, F. Oppel, B. Turowski, C. Woiciechowsky, K. Franz, T. Pietsch; ALA-Glioma Study Group "Extent of resection and survival in glioblastoma multiforme: Identification of and adjustment for bias" Neurosurgery 62(3), 564–574 (2008)
15. O. Bloch, S. J. Han, S. Cha, M. Z. Sun, M. K. Aghi, M. W. McDermott, M. S. Berger, A. T. Parsa "Impact of extent of resection for recurrent glioblastoma on overall survival: clinical article" J. Neurosurg. 117(6), 1032–1038 (2012)
16. K. L. Chaichana, E. E. Cabrera-Aldana, I. Jusue-Torres, O. Wijesekera, A. Olivi, M. Rahman, A. Quinones-Hinojosa "When Gross Total Resection of a Glioblastoma Is Possible, How Much Resection Should Be Achieved?" World Neurosurg. 82(1–2), e257–e265 (2014)
17. G. E. Keles, K. R. Lamborn, M. S. Berger "Low-grade hemispheric gliomas in adults: a critical review of extent of resection as a factor influencing outcome" J. Neurosurg. 95(5), 735–745 (2001)
18. W. Stummer, U. Pichlmeier, T. Meinel, O. D. Wiestler, F. Zanella, H. J. Reulen "Fluorescence-guided surgery with 5-aminolevulinic acid for resection of malignant glioma: a randomised controlled multicentre phase III trial" Lancet Oncol. 7(5), 392–401 (2006)
19. W. Stummer, S. Stocker, S. Wagner, H. Stepp, C. Fritsch, C. Goetz, R. Kiefmann, H. J. Reulen "Intraoperative detection of malignant gliomas by 5-aminolevulinic acid-induced porphyrin fluorescence" Neurosurgery 42(3), 516–518 (1998)
20. K. M. Hebeda, A. E. Saarnak, M. Olivo, H. J. Sterenborg, J. G. Wolbers "5-Aminolevulinic acid induced endogenous porphyrin fluorescence in 9L and C6 brain tumours and in the normal rat brain" Acta Neurochir. (Wien) 140(5), 512-513 (1998)
21. B. W. Pogue, S. Gibbs-Strauss, P. A. Valdes, K. Samkoe, D. W. Roberts, K. D. Paulsen "Review of Neurosurgical Fluorescence Imaging Methodologies" IEEE J Sel Top Quantum Electron. 16(3), 493–505 (2010)
22. P. Schucht, S. Knittel, J. Slotboom, K. Seidel, M. Murek, A. Jilch, A. Raabe, J. Beck "5-ALA complete resections go beyond MR contrast enhancement: Shift corrected volumetric analysis of the extent of resection in surgery for glioblastoma" Acta Neurochir. (Wien) 156(2), 305–312 (2014)
23. P. Schucht, J. Beck, J. Abu-Isa, L. Andereggen, M. Murek, K. Seidel, L. Stieglitz, A. Raabe "Gross total resection rates in contemporary glioblastoma surgery: results of an institutional protocol combining 5-aminolevulinic acid intraoperative fluorescence imaging and brain mapping" Neurosurgery 71(5), 926–927 (2012)
24. P. Schucht, M. Murek, A. Jilch, K. Seidel, E. Hewer, R. Wiest, A. Raabe, J. Beck "Early re-do surgery for glioblastoma is a feasible and safe strategy to achieve complete resection of enhancing tumor" PLoS One 8(11), e79846 3–9 (2013)
25. C. Nimsky, O. Ganslandt, B. von Keller, J. Romstöck, R. Fahlbusch "Intraoperative High-Field-Strength MR Imaging: Implementation and Experience in 200 Patients" Radiology 233(1), 67–78 (2004)
26. C. Senft, A. Bink, K. Franz, H. Vatter, T. Gasser, V. Seifert "Intraoperative MRI guidance and extent of resection in glioma surgery: A randomised, controlled trial" Lancet Oncol. 12(11), 997–1003 (2011)
27. P. L. Kubben, K. J. ter Meulen, O. E. M. G. Schijns, M. P. ter Laak-Poort, J. J. van Overbeeke, H. van Santbrink "Intraoperative MRI-guided resection of glioblastoma multiforme: A systematic review" Lancet Oncol. 12(11), 1062–1070 (2011)
28. A. K. Petridis, M. Anokhin, J. Vavruska, M. Mahvash, M. Scholz "The value of intraoperative sonography in low grade glioma surgery" Clin. Neurol. Neurosurg. 131, 64–68 (2015)
29. J. M. Pusterla, A. A. Malfatti-Gasperini, X. E. Puentes-Martinez, L. P. Cavalcanti, and R. G. Oliveira "Refractive index and thickness determination in Langmuir monolayers of myelin lipids" Biochim. Biophys. Acta - Biomembr. 1859, 924–930 (2017)

30. M. Axer, K. Amunts, D. Grassel, C. Palm, J. Dammers, H. Axer, U. Pietrzyk, and K. Zilles, "A novel approach to the human connectome: Ultra-high resolution mapping of fiber tracts in the brain" NeuroImage 54(2), 1091-1101 (2011)
31. M. Menzel, J. Reckfort, D. Weigand, H. Köse, K. Amunts, and M. Axer, "Diattenuation of brain tissue and its impact on 3D polarized light imaging" Biomed Opt Express 8(7), 3163-3197 (2017)
32. H. Wang, T. Akkin, C. Magnain, R. Wang, J. Dubb, W. J Kostis, M. A Yaseen, A. Cramer, S. Sakadžić, and D. Boas "Polarization sensitive optical coherence microscopy for brain imaging" Opt. Lett. 41(10), 2213-2216 (2016)
33. A. Pierangelo, A. Benali, M.-R. Antonelli, T. Novikova, P. Validire, B. Gayet, A. De Martino, "Ex-vivo characterization of human colon cancer by Mueller polarimetric imaging" Opt. Express 19(2), 1582-1593 (2011)
34. T. Novikova, A. Pierangelo, S. Manhas, A. Benali, P. Validire, B. Gayet, A. De Martino, "The origins of polarimetric image contrast between healthy and cancerous human colon tissue" Appl. Phys. Lett. 102, 241103-241106 (2013)
35. A. Pierangelo, A. Nazac, A. Benali, P. Validire, H. Cohen, T. Novikova, B. Haj Ibrahim, S. Manhas, C. Fallet, M.-R. Antonelli, A. De Martino "Polarimetric imaging of uterine cervix: a case study" Opt. Express 21(12), 14120-14130 (2013)
36. A. Pierangelo, S. Manhas, A. Benali, C. Fallet, J. L. Totobenazara M.-R. Antonelli, T. Novikova, B. Gayet, A. De Martino, P. Validire "Multi-spectral Mueller polarimetric imaging detecting residual cancer and cancer regression after neoadjuvant treatment for colorectal carcinomas" J. Biomed. Opt. 18(4), 046014 (1-9) (2013)
37. J. Vizet, J. Rehbinder, S. Deby, S. Roussel, A. Nazac, R. Soufan, C. Genestie, C. Haie-Meder, H. Fernandez, F. Moreau, A. Pierangelo "In vivo imaging of uterine cervix with a Mueller polarimetric colposcope" Sci. Rep. 7, 2471 (1-12) (2017)
38. M. Kupinski, M. Boffety, F. Goudail, R. Ossikovski, A. Pierangelo, J. Rehbinder, J. Vizet, T. Novikova "Polarimetric measurement utility for pre-cancer detection from uterine cervix specimens" Biomed. Opt. Express 9(11), 5691-5702 (2018)
39. J. Rehbinder, H. Haddad, S. Deby, B. Teig, A. Nazac, T. Novikova, A. Pierangelo, F. Moreau, "Ex vivo Mueller polarimetric imaging of the uterine cervix: a first statistical evaluation", J. Biomed. Opt. 21(7), 071113 (2016)
40. M. A. Azzam and N. M. Bashara, Ellipsometry and Polarized Light, (North-Holland: New York, Amsterdam and Oxford, 1977), pp. 148-152
41. A. De Martino, Y.-K. Kim, E. Garcia-Caurel, B. Laude, and B. Drévillon, "Optimized Mueller polarimeter with liquid crystals," Opt. Lett., 28(8), 616-618 (2003)
42. E. Compain, S. Poirier, and B. Drévillon "General and self-consistent method for the calibration of polarization modulators, polarimeters, and Mueller-matrix ellipsometers" Appl. Opt. 38(16) 3490-3502 (1999)
43. S. Y. Lu, R. A. Chipman "Interpretation of Mueller matrices based on polar decomposition" J. Opt. Soc. Am. A 13(5), 1106-1113 (1996)
44. M. Wood, N. Vurgun, M. Wallenburg, A. Vitkin "Effects of formalin fixation on tissue optical polarization properties" Phys. Med. Biol. 56(8), 115-122 (2011)
45. S. Prahl "Optical Absorption of Hemoglobin," (Portland, OR: Oregon Medical Laser Center) (1999), available at http://omlc.ogi.edu/spectra/hemoglobin/
46. A. Pierangelo, S. Manhas, A. Benali, C. Fallet, M.-R. Antonelli, T. Novikova, P. Validire, B. Gayet, A. De Martino, "Ex vivo photometric and polarimetric multilayer characterization of human healthy colon by multispectral Mueller imaging" J. Biomed. Opt. 17, 066009 (2012)
47. T. Novikova, A. Pierangelo, A. De Martino, A. Benali, P. Validire, "Polarimetric imaging for cancer diagnosis and staging", Optics and Photonics News Magazine, 26-32, (October 2012)
48. K. S. Yashin, E. B. Kiseleva, E. V. Gubarkova, A. A. Moiseev, S. S. Kuznetsov, P. A. Shilyagin, G. V. Gelikonov, I. A. Medyanik, L. Y. Kravets, A. A. Potapov, E. V. Zagaynova, N. D. Gladkova "Cross-Polarization Optical Coherence Tomography for Brain Tumor Imaging" Front. Oncol. 02 April 2019 available at https://doi.org/10.3389/fonc.2019.00201
49. L. Jiang, X. Wang, Z. Wu, H. Du, S. Wang, L. Li, N. Fang, P. Lin, J. Chen, D. Kang, S. Zhuo "Label-free imaging of brain and brain tumor specimens with combined two-photon excited fluorescence and second harmonic generation microscopy", Laser Phys. Lett. 14(10), 105401 (2017)
50. J. J. Gil and R. Ossikovski, Polarized Light and the Mueller Matrix Approach, CRC press, Taylor and Francis Group, (2016)
51. H. R. Lee, P. Li, T. S. H. Yoo, C. Lotz, F. Kai Groeber-Becker, S. Dembski, E. Garcia-Caurel, R. Ossikovski, H. Ma, T. Novikova "Digital histology with Mueller microscopy: how to mitigate an impact of tissue cut thickness fluctuations", J. Biomed. Opt. 24(7) 076004 (1-9) (2019)

**Supplementary materials**

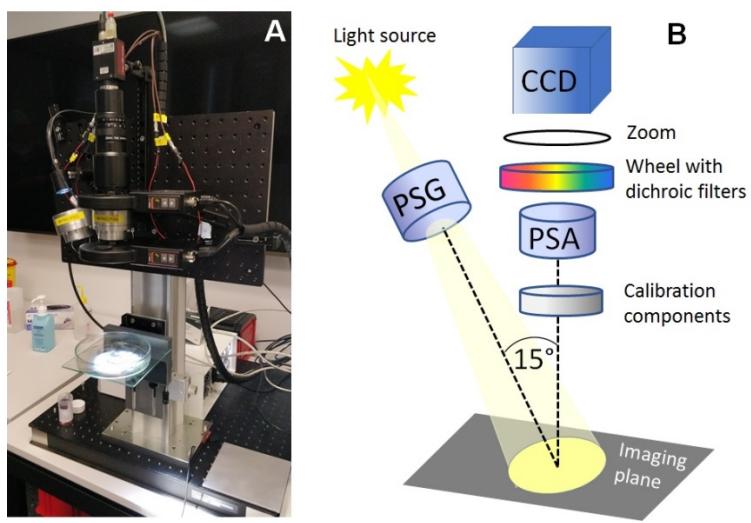

Fig. S1. Imaging Mueller polarimeter (a) photo; (b) schematic optical layout

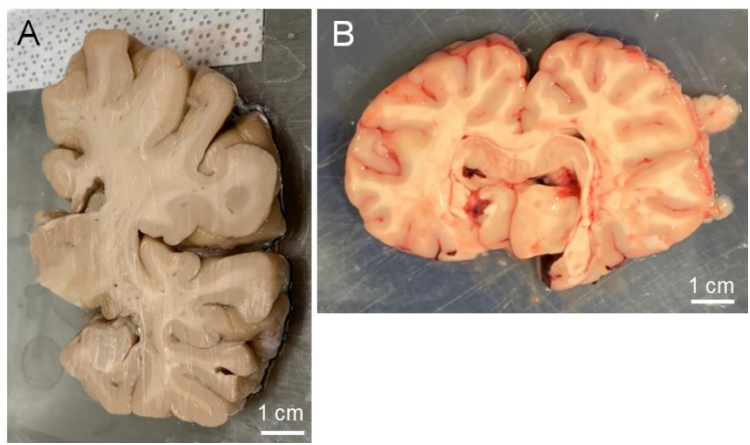

Fig. S2. Color photos of ~1 cm thick cuts in a coronal plane: A - fixed human brain (one half); B - fresh calf brain.

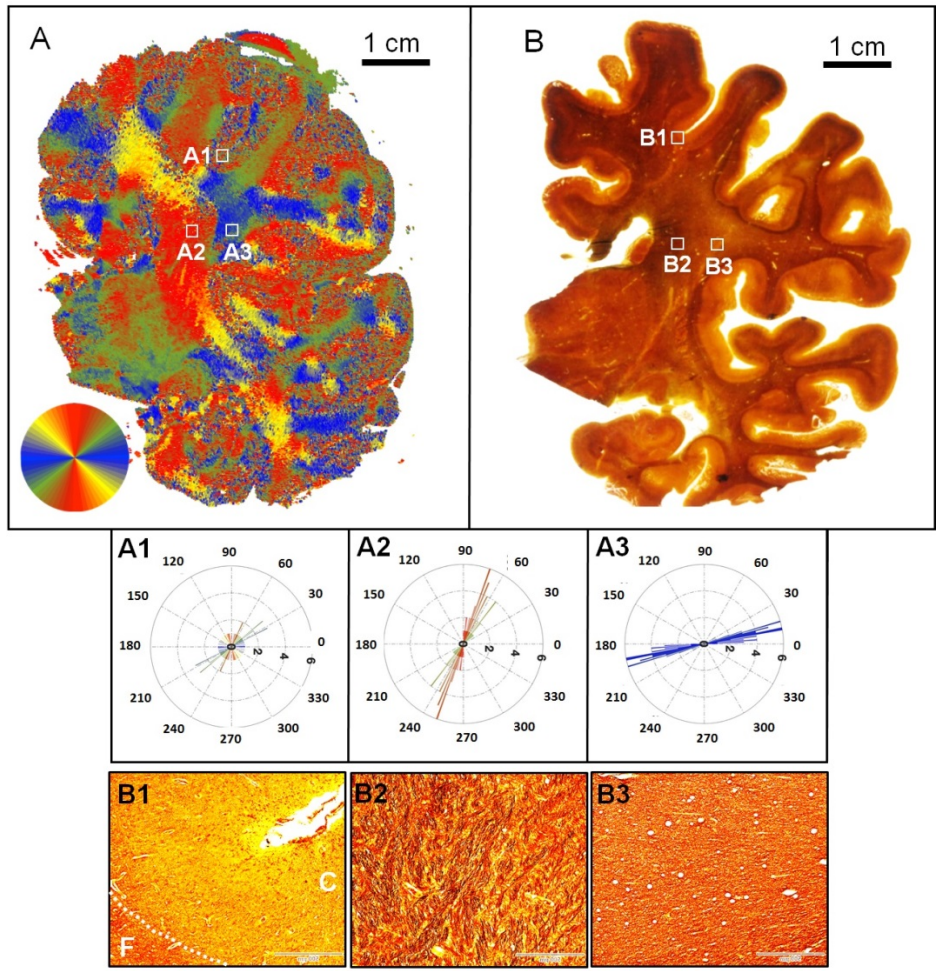

Fig. S3. Coronal section of a fixed human brain specimen: A - map of the azimuth of optical axis; B- silver-staining histology cut image; middle row – circular histograms of the azimuth of optical axis for the zones A1-A3 (9 × 9 pixels each); bottom row - the corresponding enlarged zones B1-B3 of histology cut image (C – cortex zone, F – white matter fiber tract, white dashed line represents the border between cortex and brain white matter, scale bar – 500 µm). Zones A1, B1 correspond to brain cortex; zones A2, A3, and B2, B3 represent the parts of T-junction of fiber tracks in the central part of a brain specimen.

**Tab. S1. Mean values and standard deviations of the azimuth angle distributions from the zones A1-A4 shown in Fig. 3**

| Zone | A1 | A2 | A3 | A4 |
|---|---|---|---|---|
| Mean (°) | 91.1 | 136.3 | 179.5 | 49.1 |
| Standard deviation | 10.4 | 5.9 | 11.1 | 11.8 |

**Tab. S2. Mean values and standard deviations of the azimuth angle distributions from the zones A1-A3 shown in Fig. S3.**

| Zone | A1 | A2 | A3 |
|---|---|---|---|
| Mean (°) | 13.2 | 66.5 | 12.0 |
| Standard deviation | 50.5 | 13.0 | 6.0 |